\makeatletter
\@ifundefined{@parse@version@dash}{%
\def\@parse@version#1{\@parse@version@0#1}
\def\@parse@version@#1/#2/#3#4#5\@nil{%
\@parse@version@dash#1-#2-#3#4\@nil}
\def\@parse@version@dash#1-#2-#3#4#5\@nil{%
  \if\relax#2\relax\else#1\fi#2#3#4 }
}{}
\makeatother

\documentclass[aps,prd,groupedaddress,nofootinbib,superscriptaddress,notitlepage,twocolumn]{revtex4-1}

\usepackage[normalem]{ulem}
\usepackage{graphicx}
\usepackage{dcolumn}
\usepackage{bm}
\usepackage{color}
\usepackage{array}
\usepackage{multirow}
\usepackage{amsmath}
\usepackage[colorlinks,citecolor=blue,urlcolor=blue,linkcolor=blue]{hyperref}

\newcommand{\n}{\nonumber}

\newcommand{\DM}{{\rm DM}}
\newcommand{\pccc}{\,pc\,cm$^{-3}$\,}
\newcommand{\kms}{km\,s$^{-1}$\,Mpc$^{-1}$}

\begin{document}


\title{FRB dark sirens: Measuring the Hubble constant with unlocalized fast radio bursts}


\author{Ze-Wei Zhao}
\affiliation{Key Laboratory of Cosmology and Astrophysics (Liaoning Province) \& Department of Physics, College of Sciences, Northeastern University, Shenyang 110819, China}

\author{Ji-Guo Zhang}
\affiliation{Key Laboratory of Cosmology and Astrophysics (Liaoning Province) \& Department of Physics, College of Sciences, Northeastern University, Shenyang 110819, China}

\author{Yichao Li}
\affiliation{Key Laboratory of Cosmology and Astrophysics (Liaoning Province) \& Department of Physics, College of Sciences, Northeastern University, Shenyang 110819, China}


\author{Jing-Fei Zhang}
\affiliation{Key Laboratory of Cosmology and Astrophysics (Liaoning Province) \& Department of Physics, College of Sciences, Northeastern University, Shenyang 110819, China}

\author{Xin Zhang}
\thanks{Corresponding author. \\zhangxin@mail.neu.edu.cn}
\affiliation{Key Laboratory of Cosmology and Astrophysics (Liaoning Province) \& Department of Physics, College of Sciences, Northeastern University, Shenyang 110819, China}
\affiliation{Key Laboratory of Data Analytics and Optimization for Smart Industry (Ministry of Education), Northeastern University, Shenyang 110819, China}
\affiliation{National Frontiers Science Center for Industrial Intelligence and Systems Optimization, Northeastern University, Shenyang 110819, China}

\begin{abstract}
Fast radio bursts (FRBs) can be used to measure cosmological parameters by employing the Macquart relation. However, at present, only a small number of FRB events are localized to host galaxies with known redshifts. Inspired by the dark siren method in gravitational wave cosmology, we develop a Bayesian method to statistically measure the Hubble constant using unlocalized FRBs and galaxy catalog data, which makes it possible to constrain cosmological parameters from a large number of FRB data without known redshifts, meanwhile including the real galaxy information. We assume that the probability for a galaxy to host an FRB is proportional to the luminosity of this galaxy and use the results from the IllustrisTNG simulation as the priors of FRB host galaxy parameters. Ignoring some systematic errors, we obtain the first statistical $H_0$ measurement only using twelve unlocalized FRB events combined with the big bang nucleosynthesis result, i.e., $H_0=80.4^{+24.1}_{-19.4}$ \kms\, (68\% highest-density interval). This method can also be refined to constrain other cosmological and FRB parameters. It is applicable to well-localized FRBs that still have several potential hosts.
\end{abstract}
\maketitle
\section{Introduction\label{sec:introduction}}
The Hubble tension has now been one of the most important issues in current cosmology. It shows that the values of the Hubble constant $H_0$ estimated by the early- and late-universe observations are apparently inconsistent. Actually, the disagreement of the measurements of $H_0$ has reached 4.8$\sigma$. The constraint from the Planck 2018 data of the cosmic microwave background (CMB) anisotropies \cite{Planck:2018vyg} gives $H_0=67.36\pm 0.54 $\,\kms\,, based on the $\Lambda$ cold dark matter ($\Lambda$CDM) model, and the direct measurement by the SH0ES team \cite{Riess:2021jrx} shows $H_0=73.04\pm 1.04 $\, \kms\,, using the ``distance ladder" method. In order to solve the Hubble tension, some extensions of the $\Lambda$CDM model have been proposed, but a new concordance model has not yet been widely accepted \cite{Guo:2018ans,Cai:2021weh,DiValentino:2021izs,Abdalla:2022yfr,Hu:2023jqc}.

Another possible explanation for the Hubble tension may be the unknown systematic errors in the observations. However, the current reexaminations of uncertainties are unable to fully solve the tension \cite{Abdalla:2022yfr}. Therefore, it is necessary to develop new precise cosmological probes to provide a cross-check. A key factor in developing a precise probe is that there should be a vast number of data that can be used to reduce the random error. Fast radio bursts (FRBs) are a kind of new astronomical phenomenon and it is possible for FRBs to accumulate a lot of data in the future (see Refs.~\cite{Bhandari:2021thi,Xiao:2021omr,Nicastro:2021cxs,Petroff:2021wug,Caleb:2021xqe,Zhang:2022uzl,Bailes:2022pxa} for reviews), due to their high event rate \cite{CHIMEFRB:2021srp}.

FRBs are luminous millisecond pulses detected in the radio band. When FRB photons propagate through plasma, it will interact with free electrons and generate dispersion between different frequencies. The observed high dispersion measures (DMs) greatly exceed the expectation value of the Milky Way, indicating that most FRBs are likely of extragalactic origin, except one FRB event associated with a Galactic magnetar \cite{Bochenek:2020zxn,CHIMEFRB:2020abu}. If an FRB can be localized to a unique galaxy, the redshift of this FRB can be inferred from its host galaxy \cite{Chatterjee:2017dqg,Bannister2019,Niu:2021bnl}.

Although the sources and radiation mechanisms of most FRBs have not been generally identified \cite{Wang:2016dgs,Lyubarsky:2021bai,Geng:2021apl}, one can still use them as a cosmological probe. Since DMs contain the baryonic information along the cosmological distances FRBs' photons travelled, localized FRBs can be used to constrain cosmological parameters through the Macquart relation (i.e., the DM--$z$ relation) \cite{Deng:2013aga,Zhou:2014yta,Gao:2014iva,Yang:2016zbm,Li:2017mek,Walters:2017afr,Wei:2018cgd,Liu:2019jka,Zhang:2020btn,Zhao:2020ole,Qiu:2021cww,Qiang:2021bwb,Dai:2021czy,Zhao:2021jeb,Zhu:2022mzv,Zhao:2022bpd,Wu:2022dgy,Yang:2022ftm,Zhang:2023gye}
and to constrain related parameters \cite{Li:2019klc,Cai:2019cfw,Wei:2019uhh,Wu:2020jmx,Lee:2021ppm,Gao:2022ifq,Reischke:2023gjv,Wang:2023fnn}.
An outstanding application is shown in solving the ``missing baryon" problem. Macquart \textit{et al.} \cite{Mac} used five localized FRBs to derive a cosmic baryon density constraint which is consistent with the big bang nucleosynthesis (BBN) and CMB measurements. To study the Hubble tension, the $H_0$ measurements from localized FRBs were obtained by several independent groups \cite{Hagstotz:2021jzu,Wu:2021jyk,Liu:2022bmn,James:2022dcx,Baptista:2023uqu,Fortunato:2023deh,Gao:2023izj,Wei:2023avr}, with a precision of around 10\%.

However, despite hundreds of FRB events having been detected, little more than forty FRBs are currently localized to their host galaxies \cite{Paine:2024aud}. The application of FRBs in cosmology is greatly limited because it is difficult to get the redshift information of FRBs. In order to exploit the potential of FRB data in cosmology research, it is necessary to find a way of using unlocalized FRBs to constrain cosmological parameters. In Ref.~\cite{Aggarwal:2021krc}, authors used the information from sky coordinates, galaxy fluxes, and angular sizes to estimate the probability that an FRB is associated with a candidate host.

In this work, we extend the dark siren method from gravitational wave (GW) cosmology \cite{DelPozzo:2011vcw,Chen:2017rfc,DES:2019ccw,LIGOScientific:2019zcs,Palmese:2021mjm,Song:2022siz} to the FRB field. We statistically measure the Hubble constant using the DM data of unlocalized FRBs in conjunction with galaxy catalogs.


In the dark siren method, we utilize GW events, such as binary black hole (BBH) mergers, which lack electromagnetic (EM) counterparts. By integrating these events with galaxy catalog data, we can constrain cosmological parameters by associating each GW event with its potential host galaxies. This method operates within a Bayesian framework and has been validated with mock data \cite{Gray:2019ksv}. Similarly, the Hubble constant can be constrained using unlocalized FRBs by identifying all potential host galaxies within their localization region and subsequently marginalizing over these possibilities.


Most previous studies have focused on using localized FRBs to constrain the Hubble constant \cite{Hagstotz:2021jzu,Wu:2021jyk,Liu:2022bmn,Fortunato:2023deh,Gao:2023izj,Wei:2023avr}, while another approach for unlocalized FRBs involves marginalizing the likelihood across the entire redshift range [0.01,~5] \cite{James:2022dcx,Baptista:2023uqu}. The advantage of our method is that it allows for the constraint of cosmological parameters using a large dataset of unlocalized FRBs, while also incorporating actual galaxy information (including host data) into the analysis. This enhancement potentially improves the utility of FRBs as a new cosmological probe in two significant ways: it increases the volume of data utilized and tightens the constraints derived from each unlocalized FRB event.


This paper is organized as follows. In Sec.~\ref{sec:data}, we briefly describe the  galaxy catalog and FRB data. In Sec.~\ref{sec:methodology}, the model of the likelihood and the Bayesian method are introduced. The constraints and relevant discussion are given in Sec.~\ref{sec:results}. Conclusions are given in Sec.~\ref{sec:conclusions}.

\section{Data \label{sec:data}}
\subsection{Galaxy catalog data}
The Dark Energy Spectroscopic Instrument (DESI) project is a stage IV dark energy measurement aimed at measuring the expansion of the universe and studying the physics of dark energy, by constructing a three-dimensional map of the large-scale structure of the universe. The DESI Legacy Imaging Surveys have observed the sky in three optical ($g$, $r$, and $z$) bands \cite{DESI:2018ymu}, to provide the targets for the DESI survey. The optical imaging data are collected by three independent programs (the Dark Energy Camera Legacy Survey, the Beijing-Arizona Sky Survey, and the Mayall $z$-band Legacy Survey). The 5$\sigma$ detection thresholds are 24.0, 23.4, and 22.5 AB magnitudes for the $g$, $r$, and $z$ bands, respectively, for a fiducial galaxy with an exponential light profile and half-light radius 0.45 arcsec. The optical data are combined with the infrared data at two bands, i.e., $W1$ (3.4 $\mu$m) and $W2$ (4.6 $\mu$m), observed from the Wide-field Infrared Survey Explorer (WISE) satellite \cite{Wright:2010qw}. We use the processed DESI galaxy catalog data published in Ref. \cite{Yang:2020eeb}, in which the DESI Legacy Imaging Surveys DR8 data are considered. In total, 129.35 million galaxies are selected based on the selection criteria in Ref. \cite{Yang:2020eeb}.


\subsection{FRB data}
The Australian Square Kilometre Array Pathfinder (ASKAP) project can provide precise localization for FRB events \cite{Johnston:2008hp}. The Commensal Real-time ASKAP Fast Transients (CRAFT) group \cite{CRAFT:2010dep} has performed FRB surveys in two modes: single-antenna (``Flye's Eye'', or ``FE'') mode during the ``lat50" survey, i.e. observing at Galactic latitudes $|b|\sim 50^{\circ}$ \cite{Bannister:2017sie,Shannonetal2018}, and incoherent sum (ICS) mode, incoherently adding the spectra from all antennas to localize FRBs at sub-arcsecond precision \cite{Bannister2019}. The FRBs observed by the ICS mode are mainly localized, so in this work we only use the FRB data observed by the FE mode.

\begin{figure*}[htb]
\includegraphics[width=0.7\textwidth]{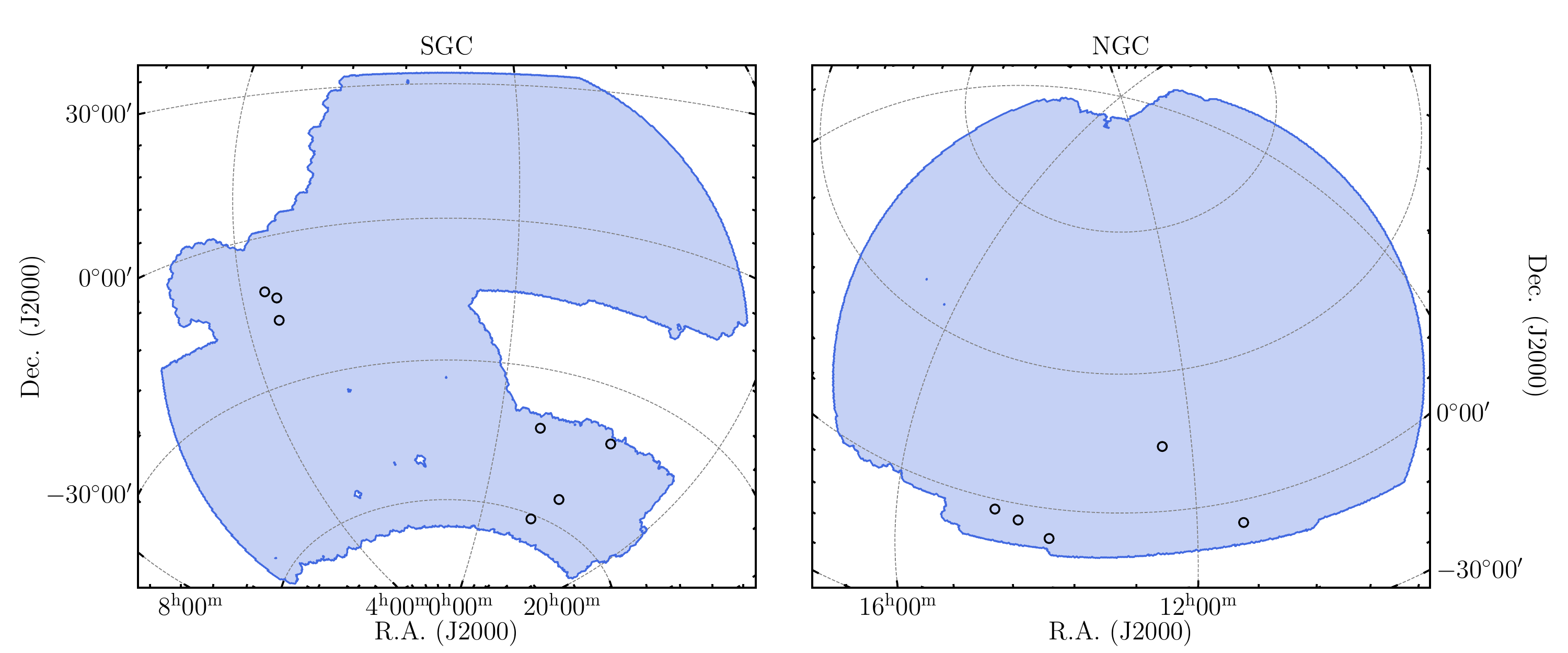}
\caption{The sky positions of the FRB data (black circles) and the footprints of the galaxy catalog (the blue area) used in this paper. The left and right panels show the data in the south Galactic cap (SGC) and north Galactic cap (NGC), respectively.}
\label{frb_footprint}
\end{figure*}

We then select the FRB events whose localization region is covered by the sky coverage of the  catalog data of the DESI Legacy Imaging Surveys. We also follow the ``gold-standard sample" criteria of FRBs for cosmological study \cite{Mac}\footnote{The first criterion \cite{Mac} only applies to localized FRBs, so we use the rest of criteria.}, and finally 12 FRB data are selected. The selected FRB data and their properties \cite{Shannonetal2018,Macquart:2018rsa,Bhandari:2019med} are listed in Table~\ref{FRBtab}, and their positions in the galaxy catalog are shown in Fig.~\ref{frb_footprint}.

\begin{table*}[!htb]
\renewcommand{\arraystretch}{1.5}
\caption{ Properties of unlocalized FRBs observed by ASKAP and the numbers of potential host galaxies ($N_{\rm gal}$) for each FRB. The names and properties of FRBs are taken from the public Transient Name Server website\footnote{https://www.wis-tns.org/}.}
\begin{tabular}{ccccc}

  \hline
  FRB event &  Right ascension & Declination  & DM (\pccc)& $N_{\rm gal}$  \\ \hline
  20170107A & 11h23m18s$\pm$10.5' & -05$^{\circ}$00'00''$\pm$10'  & 609.5& 911\\
  20170416A & 22h13m00s$\pm$15' & -10$^{\circ}$56'00''$\pm$9'  & 523.2 &375\\
  20170712A & 22h36m00s$\pm$15' & -60$^{\circ}$57'00''$\pm$10'  & 312.8 & 551\\
  20171116A & 03h31m00s$\pm$10' & -17$^{\circ}$14'00''$\pm$10'  & 618.5 &596 \\
  20171213A & 03h39m00s$\pm$30' & -10$^{\circ}$56'00''$\pm$20'  & 158.6 &4363\\
  20180119A & 03h29m18s$\pm$8' & -12$^{\circ}$44'00''$\pm$8'  &402.7 &427\\
  20180128A & 13h56m00s$\pm$20' & -06$^{\circ}$43'00''$\pm$15'  &441.4 &2425\\
  20180131A & 21h49m54s$\pm$12' & -40$^{\circ}$41'00''$\pm$8'  &657.7 &557\\
  20180212A & 14h21m00s$\pm$30' & -03$^{\circ}$35'00''$\pm$30'  & 167.5 &6811\\
  20180417A & 12h24m56s$\pm$7' & +14$^{\circ}$13'00''$\pm$7'  & 474.8 &485\\
  20180515A &  23h13m12s$\pm$7' & -42$^{\circ}$14'46''$\pm$7'  & 355.2 &270\\
  20180525A & 14h40m00s$\pm$30' & -02$^{\circ}$12'00''$\pm$6'  & 388.1 &1257\\
  \hline
\end{tabular}
\label{FRBtab}
\end{table*}

\section{Methodology\label{sec:methodology}}
\subsection{Dispersion measure distribution} \label{}
We use a DM model similar to Ref.~\cite{Mac}. The observed DM of an FRB consists of contributions from the Milky Way's interstellar medium (ISM), our Galactic halo, the cosmological distribution of the intergalactic medium (IGM) and galaxy haloes, and the FRB host galaxy,
\begin{align} \label{DMcom}
\DM  =  \DM_{\rm MW, ISM}+ \DM_{\rm MW, halo} + \DM_{\rm cosmic} + \DM_{\rm host}.
\end{align}
We use the NE2001 model \cite{Cordes:2002wz} to estimate $\DM_{\rm MW, ISM}$. The contribution from our Galactic halo with a hot, diffuse gas is still uncertain. We will discuss it below Eq.~(\ref{eqn:prob2}).

The average value of $\DM_{\rm cosmic}$ at redshift $z$ is given by the Macquart
relation \cite{Ioka:2003fr,Inoue:2003ga,Deng:2013aga},
\begin{align}\label{aveDM}
\langle\DM_{\rm cosmic}\rangle=&\int_0^z \frac{c \bar{n}_e(z^\prime) dz^\prime}{H_0 (1+z^\prime)^2 E(z')} \n\\
=&\frac{3c f_{\mathrm{d}}\Omega_\mathrm{b} H_0^2}{8\pi G m_{\mathrm{p}} H_0}\int_0^z\frac{\chi(z')(1+z')dz'}{E(z')},
\end{align}
where $\bar{n}_e$ is the mean free electron density, $\Omega_\mathrm{b}$ is the present-day baryon density parameter, $f_{\mathrm{d}}$ is the fraction of cosmic baryons in diffuse ionized gas, $\chi(z)$ represents the fraction of ionized electrons
in hydrogen and helium atoms, and $m_{\mathrm{p}}$ is the mass of a proton. The ionization fraction is
\begin{align}
\chi(z)=Y_{\rm H}\chi_\mathrm{{e,H}}(z)+\frac{1}{2}Y_{\rm He}\chi_\mathrm{{e,He}}(z).
\end{align}
where $Y_{\rm H}=3/4$ and $Y_{\rm He}=1/4$ are the mass fractions of hydrogen and helium, respectively, and $\chi_\mathrm{{e,H}}$ and $\chi_\mathrm{{e,He}}$ are the ionization fractions for hydrogen and helium, respectively. We take $\chi_\mathrm{{e,H}}=\chi_\mathrm{{e,He}}=1$, assuming both hydrogen and helium are fully ionized at $z<3$ \cite{Fan:2006dp}. For $f_{\mathrm{d}}$, we use a redshift-dependent form \cite{Mac} calculated from the public \emph{FRB} code\footnote{\textit{https://github.com/FRBs/FRB}}.

In the flat $\Lambda$CDM model, the dimensionless Hubble parameter is
\begin{align}
E(z)  =  \sqrt{\Omega_{\rm m}(1+z)^3 + (1-\Omega_{\rm m})}, \label{eq:Ez}
\end{align}
where $\Omega_{\rm m}$ is the present-day matter density parameter. We focus on the Hubble constant measurement and set $\Omega_{\rm m}=0.315$ \cite{Planck:2018vyg}. With more FRB data in future works, the influence of cosmological models and other cosmological parameters could be further studied.

Equation (\ref{aveDM}) reflects that $\langle\mathrm{DM}_{\mathrm{IGM}}\rangle$ is proportional to $\Omega_\mathrm{b}h$, with $h=H_0/(100\, {\rm km}\, {\rm s}^{-1} {\rm Mpc}^{-1})$. However, $\Omega_\mathrm{b}h^2$ is usually the directly observed quantity in other cosmological observations, so we write as $\langle\mathrm{DM}_{\mathrm{IGM}}\rangle \propto \Omega_\mathrm{b}h^2/h$, which is also a straightforward consequence of $\bar{n}_e/H_0$. Recently, a constraint $\Omega_\mathrm{b}h^2=0.02233\pm 0.00036$ is obtained only based on the big bang nucleosynthesis theory with an improved rate of deuterium burning \cite{Mossa:2020gjc}. We use its mean value as default value on baryon density, and thus the observed DM can be used to measure $H_0$.

Due to the existence of the cosmic web of filaments, voids, and other substructures, the values of $\DM_{\rm cosmic}$ have variances along different sightlines. This variation is mainly affected by the galactic feedback, based on cosmological simulations \cite{McQuinn:2013tmc}. The probability distribution of $\DM_{\rm cosmic}$ can be described by \cite{Miralda-Escude:1998adl,Mac},
\begin{align}
p_{\rm cosmic}(\Delta) = A \Delta^{-\beta} \exp\left[ - \frac{(\Delta^{-\alpha} - C_0)^2}{2 \alpha^2 \sigma_{\rm DM}^2} \right], \quad \Delta > 0,
\label{eqn:MHR}
\end{align}
where $\Delta \equiv {\rm DM}_{\rm cosmic}/\langle {\rm DM}_{\rm cosmic} \rangle$ and $A$ is a normalization factor. The parameters $\alpha=3$ and $\beta=3$ provide the best match to the models \cite{Mac}. The effective standard deviation equals approximately $F z^{-0.5}$ for $z < 1$ where the parameter $F$ describes the strength of the baryon feedback. The parameter $C_0$ can be fixed by the requirement $\langle \Delta \rangle = 1$. Combining Eqs. (\ref{aveDM})--(\ref{eqn:MHR}), we can obtain the probability distribution $p_{\rm cosmic}(\DM_{\rm cosmic}|z,H_0,F)$.


The DM contribution from host galaxy includes host galaxy halo and FRB's local environment. A log-normal distribution shows a fine fit to the results from the IllustrisTNG simulation \cite{Zhang}, and its asymmetric long tail also can account for the contribution from the gas nearby FRB sources. Hence, $\DM_{\rm host}$ can be modeled by a log-normal distribution,
{\small\begin{align}
p_{\rm host}({\rm DM}_{\rm host}^\prime) = \frac{1}{{\rm DM_{\rm host}^\prime}} \frac{1}{\sigma_{\rm host} \sqrt{2 \pi}}
\exp \left[ -\frac{(\ln {\rm DM}^\prime_{\rm host}-\mu)^2}{2 \sigma_{\rm host}^2} \right] \label{eq:phost},
\end{align}}
where ${\rm DM}_{\rm host}^\prime$ is the value of $\DM_{\rm host}$ referenced to the rest frame of the host galaxy. The mean value and variance are $e^{\mu}$ and $e^{2\mu +\sigma_{\rm host}^2} ( e^{\sigma_{\rm host}^2} - 1 )$, respectively. We apply a redshift correction ${\rm DM}_{\rm host}={\rm DM}_{\rm host}^\prime/(1+z)$ and then acquire the probability distribution $p_{\rm host}({\rm DM}_{\rm host}|z,e^\mu,\sigma_{\rm host})$.

We rewrite Eq.~(\ref{DMcom}) as
\begin{align}
\DM=\DM_{\rm MW, ISM}+\DM',
\end{align}
where $\DM'=\DM_{\rm MW, halo}+\DM_{\rm cosmic} + \DM_{\rm host}=\DM_{\rm MW, halo}+\DM_{\rm E}$ and $\DM_{\rm E}=\DM_{\rm cosmic} + \DM_{\rm host}$ is the extragalactic contribution. We ignore measurement error on DM and subtract $\DM_{\rm MW, ISM}$ from the observed DM value to obtain $\DM'$ for each FRB. Hence, the DM likelihood is
\begin{widetext}
\begin{align}
p({\rm DM}'| z,H_0,e^\mu,\sigma_{\rm host},F)  =  &\int p_{\rm halo}(\DM_{\rm MW, halo})\int_0^{\DM'-\DM_{\rm MW, halo}}  \;
p_{\rm host}({\rm DM}_{\rm host}|z,e^\mu,\sigma_{\rm host}) \n\\\; &p_{\rm cosmic}(\DM'-\DM_{\rm MW, halo} - {\rm DM}_{\rm host}|z,H_0,F) \, d{\rm DM}_{\rm host} \, d\DM_{\rm MW, halo} ,
\label{eqn:prob2}
\end{align}
where $p_{\rm host}$ and $p_{\rm cosmic}$ can be calculated from Eqs.~(\ref{aveDM})--(\ref{eq:phost}). The $p_{\rm halo}(\DM_{\rm MW, halo})$ term is the model of $\DM_{\rm MW, halo}$. The simulation \cite{Dolag:2014bca} showed that a representative halo electron contribution is about 30\pccc. By using several observations, the value of $\DM_{\rm MW, halo}$ was estimated in the range of [50, 80]\pccc \cite{prochaska2019probing}. According to these estimates, we use a Gaussian distribution $p_{\rm halo}(\DM_{\rm MW, halo})$ with the mean value of 55\pccc and standard deviation of 25\pccc to describe the distribution of $\DM_{\rm MW, halo}$ for all FRBs \cite{Wu:2021jyk}.
In principle, all the unknown parameters, i.e. $e^\mu$, $\sigma_{\rm host}$, and $F$, should be treated as free parameters and be simultaneously constrained by data. However, the purpose of this work is to demonstrate the feasibility of using unlocalized FRBs and the real galaxy catalog to measure the Hubble constant. Thus, for convenience, we use the results from other works as the priors of the parameters $e^\mu$, $\sigma_{\rm host}$, and $F$. We first assume $F=0.31$, which is a common assumption in the literature \cite{James,James:2022dcx}, based on the constraint results in Refs.~\cite{Mac,Baptista:2023uqu}.

For the FRB host galaxy parameters $e^\mu$ and $\sigma_{\rm host}$, Zhang \textit{et al.} \cite{Zhang} gave the best-fitting parameters $e^\mu\sim 32.97(1+z)^{0.84}$\pccc and $\sigma_{\rm host}\sim 1.27$ for one-off FRBs, based on the state-of-the-art IllustrisTNG simulation, which can be used as priors in the simulation-based case. Using DM model from cosmological simulations to help constrain cosmological parameters is regarded as one of the approach to avoid uncontrolled systematic errors \cite{Hollis}. They also gave that the standard deviation of $e^\mu$ in the simulation is about 15\pccc \cite{Zhang}, so we assume a Gaussian distribution $p_{\mu}(e^\mu)$ on $e^\mu$ and integrate over it. Compared to Ref.~\cite{Wu:2021jyk}, we further take the deviation of the simulation into account. Macquart \textit{et al.} \cite{Mac} also provided the median values of $e^\mu$ and $\sigma_{\rm host}$ around $68$\pccc and 0.88, respectively, using localized FRBs while treating $\Omega_\mathrm{b}h$ and $F$ as free parameters. Similarly, the observation-based case assumes a Gaussian distribution $p_{\mu}(e^\mu)$ on $e^\mu$ with a standard deviation 45\pccc \cite{Mac}. We fix the values of $\sigma_{\rm host}$ in two cases. Then Eq. (\ref{eqn:prob2}) becomes
\begin{align}
p({\rm DM}'| z,H_0)  =&   \int_{\rm 30\,pc\,cm^{-3}}^{\rm 80\,pc\,cm^{-3}}p_{\rm halo}(\DM_{\rm MW, halo})\int_{ e^\mu_{\rm min}}^{ e^\mu_{\rm max}}p_{\mu}(e^\mu) \int_0^{\DM'-\DM_{\rm MW, halo}}p_{\rm host}({\rm DM}_{\rm host}|z,e^\mu)    \n\\\;
&p_{\rm cosmic}(\DM'-\DM_{\rm MW, halo} - {\rm DM}_{\rm host}|z,H_0) \, d{\rm DM}_{\rm host}\,de^\mu \,d\DM_{\rm MW, halo} ,
\label{eqn:prob}
\end{align}
where $e^\mu_{\rm max}=18(1+z)^{0.84}$\pccc and $e^\mu_{\rm min}=48(1+z)^{0.84}$\pccc in the simulation-based case \cite{Zhang} and $e^\mu_{\rm max}=23$\pccc and $e^\mu_{\rm min}=113$\pccc in the observation-based case \cite{Mac}. This likelihood is normalized to all possible DM realizations. We again note that Eq.~(\ref{eqn:prob}) used in this paper is an approximation of Eq.~(\ref{eqn:prob2}) and we just use it in order to facilitate the calculations.

\subsection{Bayesian framework} \label{subsec:bayes}
We present an overview of the Bayesian framework to estimate $H_0$ using unlocalized FRBs. By using Bayes' theorem, the posterior of $H_0$ given the FRB data from a single detection, ${\rm d}_{\rm FRB}$, should be
\begin{align}\label{bayes}
p(H_{0}|{\rm d}_{\rm FRB})\propto p({\rm d}_{\rm FRB}|H_{0})p(H_{0}),
\end{align}
where $p({\rm d}_{\rm FRB}|H_{0})$ is the FRB likelihood, and $p(H_{0})$ is the prior on $H_0$. We use the DM likelihood, Eq.~(\ref{eqn:prob}), as a proxy of the FRB likelihood. By marginalizing over redshift and sky location $\Omega$, the FRB likelihood can be calculated by
\begin{align}\label{inteFRBEM}
	p({\rm d}_{\rm FRB}|H_{0})\propto \frac{\iint p({\rm DM}' |z,H_0)p(z,\Omega) \,d\Omega \,dz}{\beta(H_{0})} ,
\end{align}
where $\beta(H_{0})$ is a normalization term, and the probability $p(z,\Omega)$ represents an FRB occurring at redshift $z$ and sky location $\Omega$. From the galaxy catalog data, we can select the galaxies within the angular localization errors of each FRB. The photometric redshift errors of the galaxy data are about $0.01+0.015z$. But we ignore the errors and expect the final DESI data could provide precise spectroscopic redshift measurements. Then, the $p(z,\Omega)$ term can be written as the sum of the Dirac delta functions,
\begin{align} \label{pEM}
	p(z,\Omega)=\sum_{i}^{N_{\rm gal}}w_{i}\frac{R(z)}{1+z}\delta(z-z_{i})\delta(\Omega-\Omega_{i}) ,
\end{align}
where $N_{\rm gal}$ is the number of host galaxy candidates, $(z_{i}, \Omega_{i})$ represent the redshift and sky location of the $i$-th host galaxy candidate, and $w_{i}$ is the probability that the $i$-th galaxy hosts a FRB source. We consider two scenarios for this weight. The first one, called the ``equal weight" scenario, assumes equal probability for each galaxy to host an FRB source, $w_{i}=1/N_{\rm gal}$, and the second one, called the ``luminosity weight" scenario, assumes that this probability is proportional to the luminosity of this galaxy, $w_{i}\propto L_i $, meanwhile ensuring that the weights are normalized. The $R(z)$ term represents the intrinsic redshift evolution of the FRB rate, but we assume that it is redshift-independent, $R(z)={\rm constant}$. This term may also be a function of the FRB luminosity function \cite{Macquart:2018jlq,Li:2016qbl}, because the FRB luminosity could determine how far it can be detected and then affect the redshift distribution of the FRB rate. The factor $1/(1+z)$ converts the FRB rate from the source frame to the detector frame. Inserting Eq.~(\ref{pEM}) into Eq.~(\ref{inteFRBEM}), we have
\begin{align}
	p({\rm d}_{\rm FRB}|H_{0})\propto\frac{\sum^{N_{\rm gal}}_{i=1}w_{i}p({\rm DM}' |z_i,H_{0})R(z_i)/(1+z_i)}{\beta(H_{0})}.
\end{align}

The normalization term $\beta(H_{0})$ accounts for selection effects to make sure the results are unbiased. Because we have followed the ``gold-standard sample" criterion of FRBs \cite{Mac} and then selected a part of the FRB data which are enough bright to avoid the selection effect induced by detection threshold, this procedure may discard the FRBs from higher redshift, which are more likely with higher $\DM_{\rm E}$ value. So the used FRB data are below a cut-off value $\DM_{\rm cutoff}$ (in this paper $\DM_{\rm cutoff}=570$\pccc). Integrating over all possible FRB realizations below $\DM_{\rm cutoff}$ leads to
\begin{align}
\beta(H_{0})&\propto \int \int^{\DM_{\rm cutoff}}_0  p({\rm DM}_{\rm E} |z,H_{0})p(z) \,d {\rm DM}_{\rm E}\,d z  \n\\
&\propto \sum^{N_{\rm gal}}_{i=1}\int^{\DM_{\rm cutoff}}_0p({\rm DM}_{\rm E} |z_i,H_{0})w_{i}\frac{R(z_i)}{1+z_i}\,d {\rm DM}_{\rm E}
, \label{eq:Ft}
\end{align}
where $p(z)$ is identical to Eq.~(\ref{pEM}). 

For multiple FRB events $\{{\rm d}_{{\rm FRB},j}\}$, we assume that they are independent of each other and rewrite Eq.~(\ref{bayes}) as
\begin{align}
	p(H_{0}|\{{\rm d}_{{\rm FRB},j}\})\propto p(H_{0}) \prod_j p({\rm d}_{{\rm FRB},j}|H_{0}). \label{eqn:pos}
\end{align}
We take the prior on $H_0$ to be uniform over the range [20,~140] $\mathrm{km}\, \mathrm{s}^{-1}\,\mathrm{Mpc}^{-1}$. We use the Markov chain Monte Carlo method to sample the posterior and estimate constraints, which is implemented by the {\tt emcee} package \cite{Foreman-Mackey:2012any} with 32 walkers of 20000 samples after discarding 200 burn-in steps.

\end{widetext}

\section{Results and discussion\label{sec:results}}
\begin{figure*}[htb]
\includegraphics[width=0.7\textwidth]{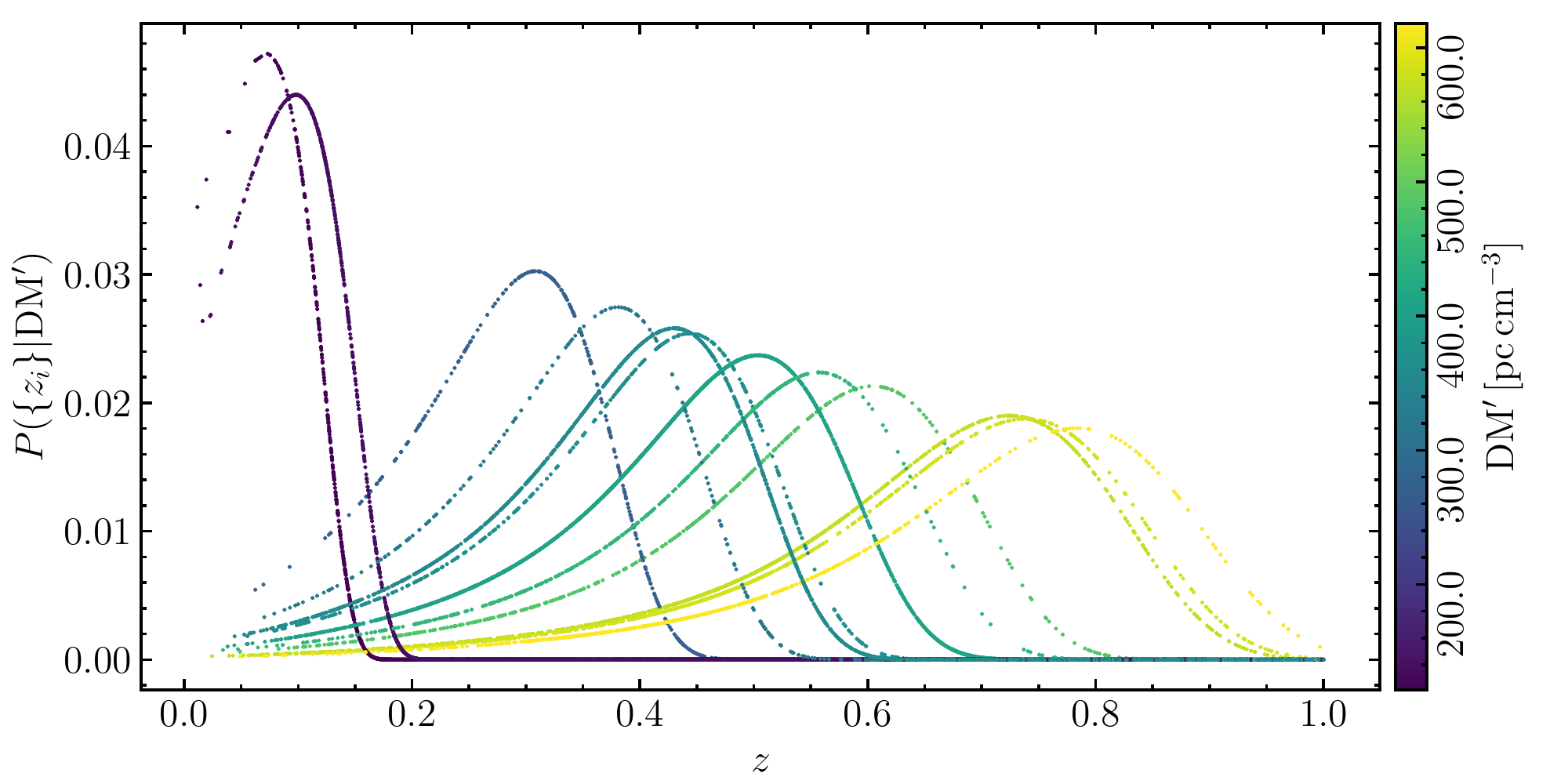}
\caption{The discrete probabilities $p(\{z_i\}|\DM')$ calculated at the redshifts of potential hosts, $\{z_i\}$, for each FRB event in the ``equal weight" scenario. The ${\rm DM}'$ value of FRB events (from low to high) are indicated by the colors of lines (from purple to yellow). Note that we assume $\DM_{\rm MW, halo}=50$\pccc, $H_0=70\,$\kms, $e^\mu=50$\pccc, and $\sigma_{\rm host}=1$ as an example.}
\label{like}
\end{figure*}

The core of the analysis is the probability $p(z_i|\DM')\propto p(\DM'|z_i)p(z_i)$ calculated at the redshift of the $i$-th potential host, $z_i$. As an example, the discrete probability $p(z_i|\DM')$ in the ``equal weight" scenario is shown in Fig.~\ref{like}, with the assumption $\DM_{\rm MW, halo}=50$\pccc, $H_0=70\,$\kms, $e^\mu=50$\pccc, and $\sigma_{\rm host}=1$. It demonstrates that the galaxies above the maximum redshift of the galaxy catalog ($z> 1$) could almost contribute little to the total likelihood.

\begin{figure*}[htb]
\includegraphics[width=0.7\textwidth]{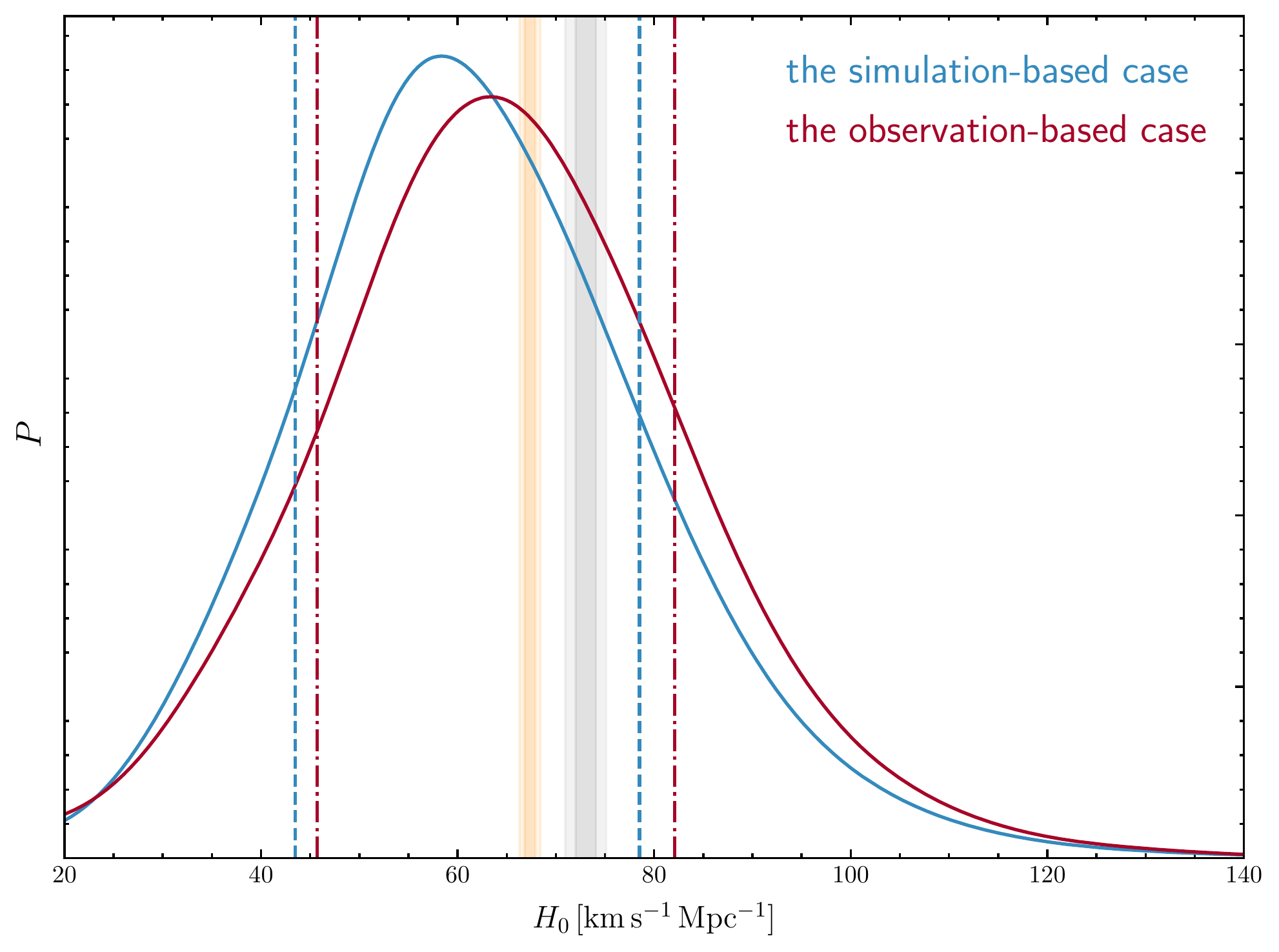}
\caption{
{The posteriors for $H_0$ from 12 unlocalized ASKAP FRBs in the ``equal weight" scenario. The constraints for the simulation-based case and the observation-based case are shown in blue and red solid lines, respectively, and their 1$\sigma$ credible intervals are shown in vertical dotted lines. The constraints from Planck \cite{Planck:2018vyg} and SH0ES \cite{Riess:2021jrx} are also shown in orange and grey regions, respectively.}}
\label{H0fig}
\end{figure*}

In Fig.~\ref{H0fig}, we show the constraints on $H_0$ assuming different host galaxy parameters from 12 unlocalized ASKAP FRBs in the ``equal weight" scenario. The maximum a posteriori (MAP) values with the minimal 68.3\% credible intervals are $H_0=58.7^{+19.8}_{-15.2}$ \kms\, in the simulation-based case and $H_0=62.3^{+19.8}_{-16.6}$ \kms\, in the observation-based case. Two estimates are basically consistent, and the difference can be regarded as a systematic error until we know more about the host galaxy parameters. The results of Planck and SH0ES are both within the 1$\sigma$ credible intervals of these two estimates. At current precision, our measurement cannot help resolve the Hubble tension.


\begin{figure*}[htb]
\includegraphics[width=0.7\textwidth]{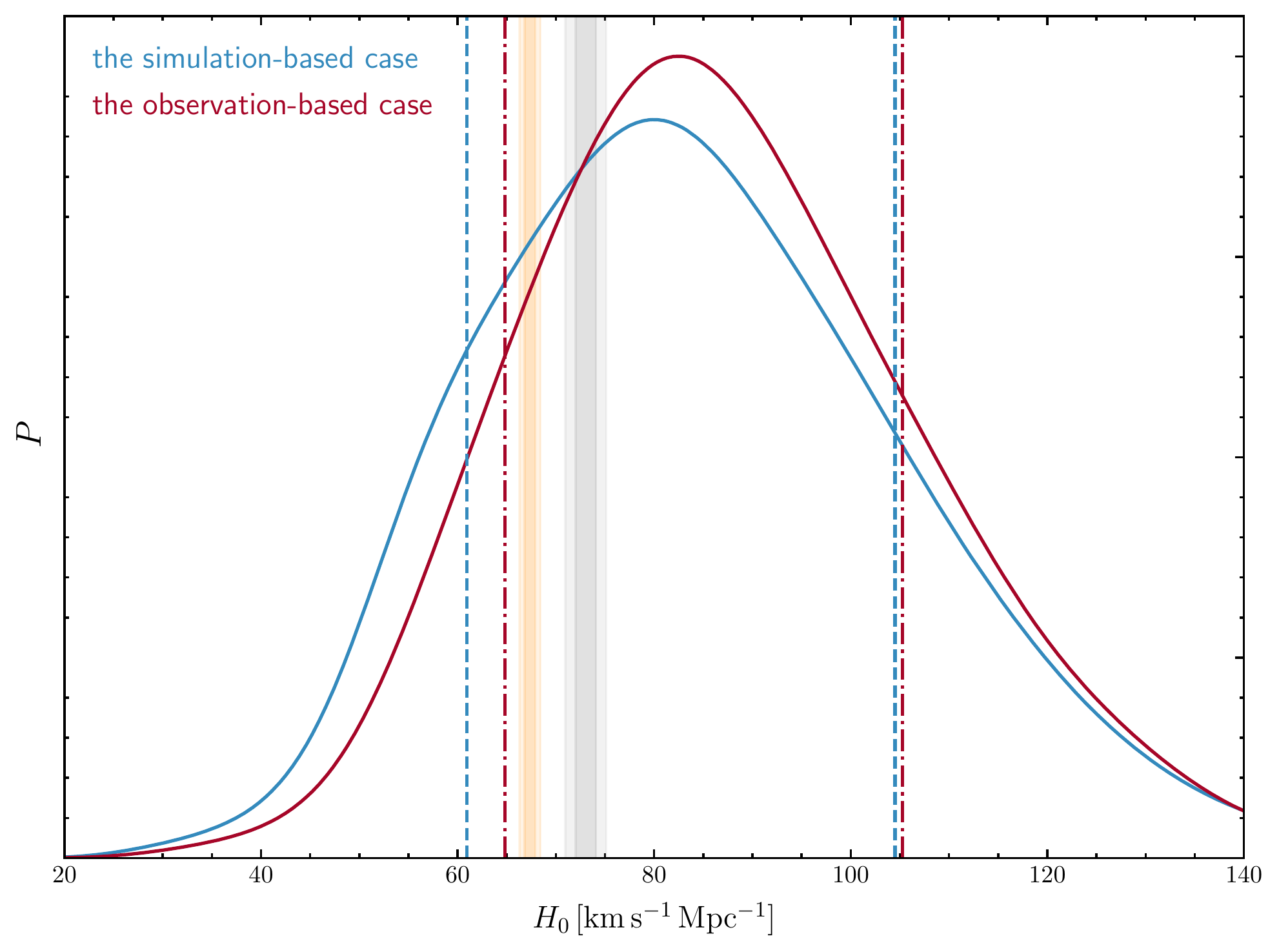}
\caption{
{The posteriors for $H_0$ from 12 unlocalized ASKAP FRBs in the ``luminosity weight" scenario. The constraints for the simulation-based case and the observation-based case are shown in blue and red solid lines, respectively, and their 1$\sigma$ credible intervals are shown in vertical dotted lines. The constraints from Planck \cite{Planck:2018vyg} and SH0ES \cite{Riess:2021jrx} are also shown in orange and grey regions, respectively.}}
\label{H0lumfig}
\end{figure*}

It should be noted that actually, above $z\sim 0.5$, the galaxy catalog is incomplete \cite{Yang:2020eeb}, thus some high-redshift galaxies are absent in our analysis. This leads to the constraints lower compared to the reality (i.e., with a complete galaxy catalog), because of the positive correlation between $z$ and $H_0$ [Eq.~(\ref{aveDM})]. The ``luminosity weight" scenario is a reasonable way to reduce this effect. Because the absent high-redshift galaxies are with low luminosity, they would only contribute a little to the likelihood and final results. We should admit that the incompleteness of galaxy catalog needs a full analysis, which can refer to the calculations in the dark siren method \cite{Gray:2019ksv}, but this calculation is extremely time-consuming for FRBs due to the multiple integral in the DM likelihood, Eq.~(\ref{eqn:prob}).


\begin{figure*}
\includegraphics[width=0.7\textwidth]{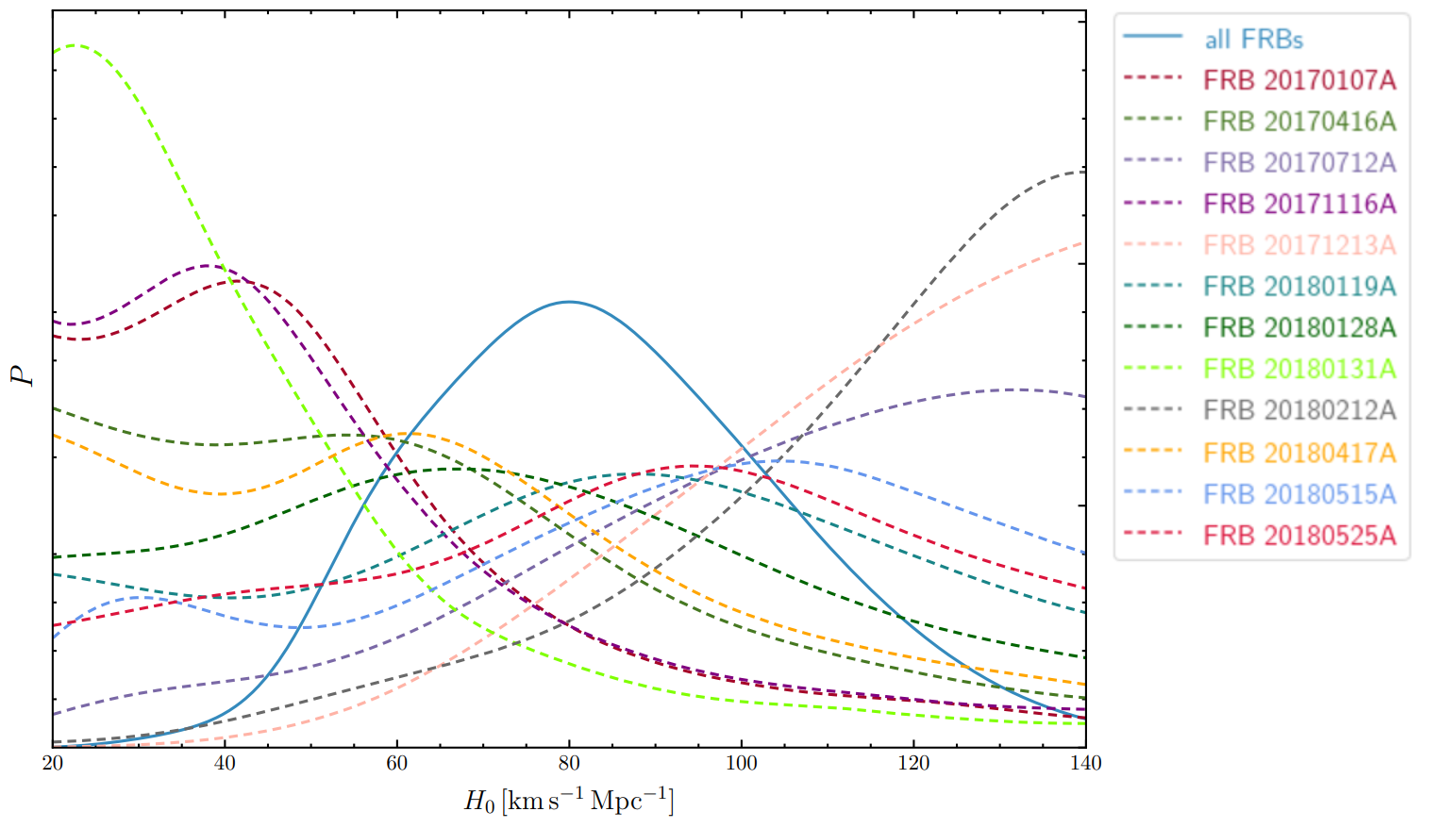}
\caption{
The posteriors for $H_0$ from each unlocalized ASKAP FRB. The constraints for individual FRB are shown in dotted lines, and the combined constraint is shown in solid line.}
\label{everyFRB}
\end{figure*}

Figure~\ref{H0lumfig} is the same as Fig.~\ref{H0fig} but in the ``luminosity weight" scenario. We can clearly see that the $H_0$ constraints are obviously increased in this scenario. The simulation-based and observation-based cases provide constraints, $H_0=80.4^{+24.1}_{-19.4}$ \kms\, and $H_0=81.9^{+23.4}_{-17.1}$ \kms, respectively. In Fig.~\ref{everyFRB}, we plot the posterior distributions of $H_0$ for each FRB event in the simulation-based case and the ``luminosity weight" scenario. We can see that although most of the single events only provide very weak constraints, the combined constraint can provide an informative constraint around 80 \kms.

We compare our $H_0$ constraints with the ones in the literature. The constraints, $H_0=62.3\pm 9.1$ \kms\, from 9 localized FRBs \cite{Hagstotz:2021jzu}, $H_0=68.81^{+4.99}_{-4.33}$ \kms\, from 18 localized FRBs \cite{Wu:2021jyk}, $H_0=70.60\pm2.11$ \kms\, from a cosmological-model-independent method \cite{Liu:2022bmn}, $H_0=69.4\pm4.7$ \kms\, from 23 well-localized FRBs \cite{Fortunato:2023deh}, and $H_0=65.5^{+6.4}_{-5.4}$ \kms\, from Supernova Ia combined with 18 localized FRBs \cite{Gao:2023izj} are all lower than our results in the ``luminosity weight" scenario, but still consistent with our results at the 1$\sigma$ credible level, due to the current loose constraints. Another work setting a flat prior on $\DM_{\rm MW, halo}$ provided $H_0=95.8^{+7.8}_{-9.2}$ \kms \cite{Wei:2023avr}, which is obviously higher than the estimates above but consistent with our results. The other two works \cite{James:2022dcx,Baptista:2023uqu} using both localized and unlocalized FRBs gave $H_0=73^{+12}_{-8}$ \kms\, and $H_0=85.3^{+9.4}_{-8.1}$ \kms, respectively,  which are closer to our MAP value, 80 \kms, of the ``luminosity weight" constraint. Note that their analysis just marginalized the likelihood over a broad redshift range for unlocalized FRBs, but without the real galaxy information.


We also compare the relative errors of our results with the one using dark sirens in GW astronomy. 46 GW events combined with the GLADE+ galaxy catalog data could give about 19\% precision constraint on $H_0$ \cite{LIGOScientific:2021aug}, yet our method achieves about 27\% precision constraint using 12 FRB events. By approximating a Gaussian distribution and scaling our result as $1/\sqrt{N}$ behaviour, where $N$ is the number of FRB events, we see that our method can provide similar constraint ability to the dark siren method.

The 12 FRB events account for about a half of the total FRB events observed by ASKAP FE mode, because the catalog footprint covers about a half of the total sky. Applying this proportion to the total current FRB data, we would have about 400 events to perform this analysis. Again scaling our result as $1/\sqrt{N}$ behaviour, the constraint on $H_0$ is able to reach 8\% precision using those 400 unlocalized FRBs. However, at that point, the impacts of the host galaxy parameters, the choice of priors and cosmological models, and other systematic errors would become more important and should be carefully treated. We will study the systematic errors of this method in the next paper.

We expect that this method can be further developed to measure other cosmological parameters and FRB parameters \cite{Li:2023zro}, benefiting by the high detection rates of FRBs. Actually, even for some of the well-localized FRBs (such as being localized to arcsecond), it is still possible more than one single obvious host galaxy in the FRB localization region. Therefore, cosmological parameter estimations still have to be marginalized over multiple potential hosts for these FRBs.

\section{Conclusions\label{sec:conclusions}}
In this paper, we use the statistical galaxy catalog method similar to the dark siren method in GW cosmology and obtain the measurement of the Hubble constant $H_0$ using twelve unlocalized ASKAP FRB data as an example. We find all potential host galaxies in the FRB localization region and then marginalize the FRB likelihood over them. For the weights on the probability for a galaxy to host an FRB source, two different scenarios are considered, i.e. the ``equal weight" scenario and the ``luminosity weight" scenario. We also use the results from the IllustrisTNG simulation and the constraints from the localized FRBs as two priors of the ${\rm DM}_{\rm host}$ model, denoted by the simulation-based and observation-based cases, respectively. In the ``equal weight" scenario, the constraints are $H_0=58.7^{+19.8}_{-15.2}$ \kms\, in the simulation-based case and $H_0=62.3^{+19.8}_{-16.6}$ \kms\, in the observation-based case, combined with the BBN result on $\Omega_{\rm b}h^2$. To reduce the effect of the galaxy catalog incompleteness, we obtain the constraints in the ``luminosity weight" scenario, $H_0=80.4^{+24.1}_{-19.4}$ \kms\, and $H_0=81.9^{+23.4}_{-17.1}$ \kms\,, in the simulation-based and observation-based cases, respectively. These results are very rough estimates, but serve to demonstrate the feasibility of this method.

We should point out one disadvantage of this work that we used the priors for the host galaxy parameters, i.e. $e^\mu$ and $\sigma_{\rm host}$. In principle, they should be regarded as free parameters. The results obtained by fixing these parameters may be biased \cite{James:2022dcx}. Actually, this problem also exists in early study of the dark siren method in GW cosmology, in which the underlying source mass distribution is fixed, and the systematic errors are studied by a series following papers. Nonetheless, the deviation between the cases of different host galaxy parameters could be regarded as a systematic error, until more accurate information about FRB hosts is known. It should be noted that other emerging $H_0$ measurements (such as standard sirens \cite{Schutz:1986gp,Holz:2005df,LIGOScientific:2017adf,LIGOScientific:2017vwq,LIGOScientific:2017ync,Zhao:2010sz,Vitale:2018wlg,Wang:2018lun,Jin:2020hmc,Belgacem:2019tbw,Howlett:2019mdh,Zhang:2019loq,Ezquiaga:2022zkx,Wang:2019tto,Zhao:2019gyk,Wang:2021srv,Jin:2021pcv,Yu:2021nvx,Jin:2023sfc,Zhu:2023jti,Han:2023exn,Li:2023gtu}) also have different systematic errors, e.g., systematic errors of bright sirens from the viewing angle of binary neutron stars \cite{Chen:2020dyt}, of dark sirens from the assumed BBH mass distribution \cite{LIGOScientific:2021aug}, of localized FRBs from tentative host galaxy \cite{Mac,Seebeck:2021szj}. So it is still important to develop new $H_0$ measurements to provide a cross-check for each other.


\begin{acknowledgments}
We are very grateful to Jia-Ming Zou, J. X. Prochaska, Zheng-Xiang Li, and Yougang Wang for fruitful discussions.
This work was supported by the National SKA Program of China (Grants Nos. 2022SKA0110200 and 2022SKA0110203),
the National Natural Science Foundation of China (Grants Nos. 11975072, 11875102, and 11835009),
the science research grants from the China Manned Space Project (Grant No. CMS-CSST-2021-B01),
the Liaoning Revitalization Talents Program (Grant No. XLYC1905011),
the National Program for Support of Top-Notch Young Professionals (Grant No. W02070050),
and the National 111 Project of China (Grant No. B16009).

\end{acknowledgments}

\bibliography{FRBH0ref}



\end{document}